# Concepts of Ferrovalley Material and Anomalous Valley Hall Effect

Wen-Yi Tong,[1] Shi-Jing Gong,[1] Xiangang Wan,[2,3] and Chun-Gang Duan[1,4,*]

[1]*Key Laboratory of Polar Materials and Devices, Ministry of Education, East China Normal University, Shanghai, Shanghai 200241, China*
[2]*National Laboratory of Solid State Microstructures and Department of Physics, Nanjing University, Nanjing, Jiangsu 210093, China*
[3]*Collaborative Innovation Center of Advanced Microstructures, Nanjing University, Nanjing, Jiangsu 210093, China*
[4]*Collaborative Innovation Center of Extreme Optics, Shanxi University, Taiyuan, Shanxi 030006, China*
(Date: 20 April 2016)

Valleytronics rooted in the valley degree of freedom is of both theoretical and technological importance as it offers additional opportunities for information storage and electronic, magnetic and optical switches. In analogy to ferroelectric materials with spontaneous charge polarization in electronics, as well as ferromagnetic materials with spontaneous spin polarization in spintronics, here we introduce a new member of ferroic-family, i.e. a *ferrovalley* material with spontaneous valley polarization. Combining a two-band **k·p** model with first-principles calculations, we show that 2H-VSe$_2$ monolayer, where the spin-orbit coupling coexists with the *intrinsic* exchange interaction of transition-metal-*d* electrons, is such a room-temperature ferrovalley material. We further predict that such system could demonstrate many distinctive properties, for example, chirality-dependent optical band gap and more interestingly, *anomalous valley Hall effect*. On account of the latter, a series of functional devices based on ferrovalley materials, such as valley-based nonvolatile random access memory, valley filter, are contemplated for valleytronic applications.

With the celebrated discovery of graphene [1], the concept of valleytronics based on graphene-related materials (GRMs) with honeycomb lattice symmetry has attracted immense attention [2-5]. Similar to charge and spin of electrons in electronics and spintronics, the valley degree of freedom in the field of valleytronics, corresponding to degenerate but unequivalent $K_+$ and $K_-$ points (so called valleys) at the corners of the two-dimensional (2D) hexagonal Brillouin zone, constitutes the binary states. This leads to a great deal of unconventional phenomena and possibilities for practical applications, especially in information processing industry.

Among GRMs, monolayers of 2H-phase transition-metal dichalcogenides (TMDs) [6-9] are the "star of outlook" with unique potential for utilizing and manipulating of valley index effectively. At variance with graphene, the space inversion symmetry for these 2D materials are explicitly broken, which gives rise to the existence of the valley Hall effect [4], as well as the valley-dependent optical selection rules [10]. Particularly, the noncentrosymmetry together with intrinsic spin-orbit coupling (SOC) originated from the *d*-orbitals of heavy transition metals [11] induce strong coupled spin and valley degree of freedom, making them as a promising platform for the study of the fundamental physics in spintronics, valleytronics and crossing areas.

In analogy with paraelectric and paramagnetic materials, the pristine TMDs monolayers are not suitable for long-term storing information. In this regard, the major challenge in valleytronics is to break the degeneracy between the two prominent $K_+$ and $K_-$ valleys, i.e. to achieve the valley polarization. At present stage, the principal mechanism invoked in the context is circularly polarized optical excitation [12-14]. However, as a dynamic process, optical pumping merely changes the chemical potential in two valleys. It does not meet the requirement of robust manipulation. In the presence of an external electric field, electron-electron interaction-driven broken time-reversal symmetry is expected to be an effective way to induce valley polarization [3]. Another typical strategy through an external magnetic field [15-18], as it turns out, indeed lift the valley degeneracy energetically. Unfortunately, the extreme field strength for a sizable valley splitting is unaccessible in practical use. The approach based on valley-selective optical Stark effect appears to suffer from a similar problem due to the huge required amplitude of oscillating electric field [19]. Another limitation of the attempts mentioned above and some others, such as strain engineering [20, 21], lies in the volatility. When the applied external fields including force, electric, magnetic and optical ones remove, the valleys locked by time-reversal symmetry are still degenerate, stabilizing the system in the initial paravalley state. For the purpose of applying in next-generation electronic products with nonvolatility, scheme by means of magnetic doping [22-26] appeared as an alternative approach. In consideration of the electronic transports suffering from impurity scattering, a more intelligent way, i.e. using the magnetic proximity effect [27, 28] is proposed very recently. Although there exists giant and tunable valley degeneracy splitting in MoTe$_2$ induced by EuO, it is still an external method. To explore *intrinsic* valley polarization in TMDs is thus highly desirable.

In this Letter, we rebuild the Hamiltonian for the classical monolayers of TMDs, and point out that the coexistence of the SOC effect and exchange interaction of localized *d*-electrons is the sufficient condition for spontaneous valley polarization. In addition to ferroelectric and ferromagnetic materials that have been routinely explored, we therefore for the first time unveil a *ferrovalley* material, as a new ferroic-family member. We predict that intriguing phenomena like *anomalous valley Hall effect* could occur in such system.

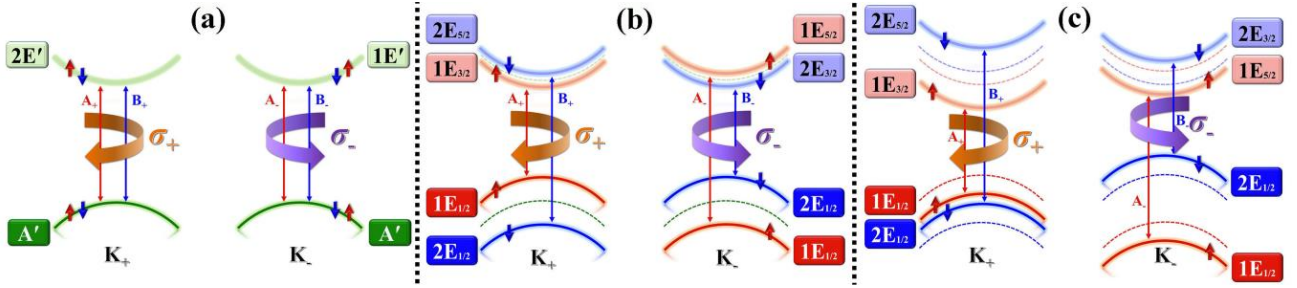

FIG. 1 (color online). The schematic band structures at valleys $K_+$ and $K_-$ of representative 2H-phase TMDs monolayers (a) without SOC effect, (b) with SOC effect, and (c) with SOC effect and a positive exchange field, i.e. valley polarized case. The IRs of states have been labelled using the Mulliken notations. The allowed interband transitions excited by circularly-polarized light near the band edges have been plotted as $A_+$, $B_+$, $A_-$ and $B_-$. $\sigma_+$ and $\sigma_-$ represent the left-handed and right-handed radiation, respectively. Noted that the band structures are referenced to the one of monolayer $MoS_2$. The spin splitting of UB induced by SOC effect has the opposite sign to the one of LB. In addition, the effective exchange splitting in the band edge of UB considered here is slightly larger than that of LB.

It is known that for representative monolayers of 2H-phase TMDs with trigonal prismatic coordination ($D_{3h}$) [29, 30], such as $MoS_2$, the direct band gaps are located at valleys $K_+$ and $K_-$ with $C_{3h}$ point group symmetry. The bottom of the conduction band (CB) dominantly consists from $d_{z^2}$ orbitals on transition metal involving with a minor contribution from the $p$-orbitals of chalcogens. At the top of the valance band (VB), there exists mainly hybridization between $d_{x^2-y^2}$ and $d_{xy}$ states of cation to interact with $p_x$ and $p_y$ states on anions. A two-band $\mathbf{k}\cdot\mathbf{p}$ model neglecting $p$-orbitals on the chalcogen with $|\psi_u^\tau\rangle = |d_{z^2}\rangle$ and $|\psi_l^\tau\rangle = (|d_{x^2-y^2}\rangle + i\tau|d_{xy}\rangle)/\sqrt{2}$ ($\tau = \pm 1$ denotes the valley index) as basis functions can be used to describe the electronic properties near the Dirac points $K_\pm$ [4, 9]. Here, subscripts $u$ (upper band, UB) and $l$ (lower band, LB), instead of CB and VB, are adopted to describe the valley states. In order to violate the time inversion symmetry and induce the valley polarization, additional term $H_{ex}(\mathbf{k})$ is introduced to the effective Hamiltonian. We then construct the total Hamiltonian as follows:

$$H(\mathbf{k}) = I_2 \otimes H_0(\mathbf{k}) + H_{SOC}(\mathbf{k}) + H_{ex}(\mathbf{k}). \quad (1)$$

To reproduce the anisotropic dispersion and more importantly the electron-hole asymmetry, the first term with up to second-nearest-neighbor hopping is given by [31, 32]:

$$H_0(\mathbf{k}) = \begin{bmatrix} \frac{\Delta}{2} + \varepsilon + t_{11}'(q_x^2 + q_y^2) & t_{12}(\tau q_x - iq_y) + t_{12}'(\tau q_x + iq_y)^2 \\ t_{12}(\tau q_x + iq_y) + t_{12}'(\tau q_x - iq_y)^2 & -\frac{\Delta}{2} + \varepsilon + t_{22}'(q_x^2 + q_y^2) \end{bmatrix}, \quad (2)$$

in which $\Delta$ is band gap at the valleys ($K_\pm$), $\varepsilon$ is a correction energy bound up with the Fermi energy, $t_{12}$ is the effective nearest neighbor hopping integral, $t_{11}'$, $t_{12}'$ and $t_{22}'$ are parameters related to the second-nearest-neighbor hopping, and $\mathbf{q} = \mathbf{k} - \mathbf{K}$ is the momentum vector. $I_2$ is the $2 \times 2$ identity matrix.

The second term, i.e. the SOC term can be written as:

$$H_{SOC}(\mathbf{k}) = \frac{\tau\lambda}{2}\begin{bmatrix} L_z & L_x - iL_y \\ L_x + iL_y & -L_z \end{bmatrix} + H_{SOC}' \quad (3)$$

Here, $L_x$, $L_y$, $L_z$ are the $2 \times 2$ matrix for $x$, $y$, $z$ components of the orbital angular momentum. The perturbation correction $H_{SOC}'$, as a valley-dependent $4 \times 4$ matrix, is applied here to incorporate the contributions from $p$-orbitals of anions and the remote $d_{xz}$ and $d_{yz}$ characters on transition metal [31]. The SOC effect directly causes the spin splitting at the bottom of the UB (the top of the LB). We label it as $2\lambda_u$ ($2\lambda_l$), defined by the energy difference $E_{u(l)\uparrow} - E_{u(l)\downarrow}$ at the $K_+$ point.

The most crucial term we imported originates from the *intrinsic* exchange interaction of transition-metal $d$-electrons:

$$H_{ex}(\mathbf{k}) = \sigma_z \otimes \begin{bmatrix} -m_u & 0 \\ 0 & -m_l \end{bmatrix}, \quad (4)$$

where $\sigma_z$ is the Pauli matrix, and $m_u$ ($m_l$) = $E_{u(l)\downarrow} - E_{u(l)\uparrow}$ represents the effective exchange splitting in the band edge of UB (LB). The exchange interaction, equivalent to an *intrinsic* magnetic field, tends to split the spin-majority and spin-minority states. Combining the valley-independent exchange interaction with valley-dependent SOC effect, the valley polarization is therefore feasible.

According to the total Hamiltonian, the band structures near the valleys $K_\pm$ of classical TMDs monolayers are easily deduced and schematically drawn in Fig. 1, in which the Fermi level is located at the gap between UB and LB. In spite of the identical occupations, the symmetry between points $K_+$ and $K_-$ is quite different. Previous work [14] has successfully proposed the chiral absorptivities in valleytronic materials based on conservation of overall azimuthal quantum number. Here, using group theory analysis, we systematically explore the valley-dependent optical selection rules, as well as the impact of valley polarization on optical properties.

As shown in Fig. 1a with absence of the SOC effect, the irreducible representations (IRs) at $K_+$ are A′ and 2E′ for LB and UB, respectively. While for $K_-$, the one for the bottom of the UB changes as 1E′. The IRs of states are labelled in Mülliken notations. Due to the Wonderful Orthogonality Theorem, the electric-dipole transition is forbidden unless the reduction of the product representation between the IRs of initial state and the incident radiation contains the representation of the final state. When the incident light is left-handed (right-handed) circularly polarized with 2E′

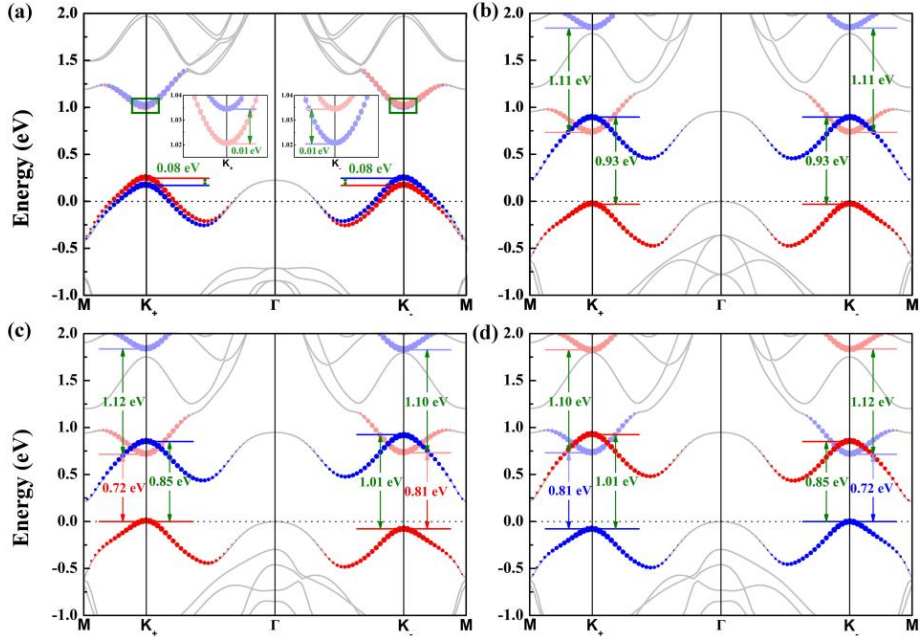

FIG. 2 (color online). The band structures of 2H-VSe$_2$ monolayer (a) with SOC effect but without ferromagnetism, and (b) with magnetic moment but without SOC effect. (c) is the real case including both the magnetism and SOC effect. (d) is same as (c) but with opposite magnetic moment. The insets in (a) amplify the spin splitting at the bottom of the CB. The radius of dots is proportional to its population in corresponding state: red and blue ones for spin-up and spin-down components of $d_{x^2-y^2}$ and $d_{xy}$ orbitals on cation-V, and light red and light blue symbols represent spin-up and spin-down states for $d_{z^2}$ characters. The Fermi level $E_F$ is set to zero in each cases.

(1E′) symmetry, we note that A′ ⊗ 2E′ (1E′) = 2E′ (1E′). Apparently, the optical absorption at K$_+$ (K$_-$) could only be excited by the left-handed (right-handed) light, which implies the valley-dependent optical selection rules. At present, the optical band gaps ($E_g^{opt}$) are identical, i.e. $E_g^{opt}$(A$_+$) = $E_g^{opt}$(B$_+$) = $E_g^{opt}$(A$_-$) = $E_g^{opt}$(B$_-$) = Δ.

When the SOC effect is taken into account, the symmetry for valleys has to be interpreted by the double group $C_{3h}^D$. The identical representation A′ degenerates to 1E$_{1/2}$ and 2E$_{1/2}$ for spin-up and spin-down components. Meanwhile, 2E′ and 1E′ change to 1E$_{3/2}$ (spin-up), 2E$_{5/2}$ (spin-down) and 1E$_{5/2}$ (spin-up), 2E$_{3/2}$ (spin-down), accordingly. By applying direct product between IRs of the ground state and circularly polarized light, the allowed interband transitions can be easily obtained, as labelled in Fig. 1b. Remarkably, the chirality is locked in each valley. The valley-dependent SOC effect splits the previously degenerated A$_+$ (A$_-$) and B$_+$ (B$_-$), and makes the $E_g^{opt}$ in K$_+$ and K$_-$ stemming from different spin states. Nevertheless, they still bear the same value ($E_g^{opt}$(A$_+$) = $E_g^{opt}$(B$_-$) = Δ − $\lambda_l$ + $\lambda_u$), owing to the protection by time-reversal symmetry.

The existence of intrinsic exchange interaction (Fig. 1c) breaks the inversion symmetry and decoupled the energetically degenerated valleys, clearly elucidating the occurrence of valley polarization. It is interesting to point out that $E_g^{opt}$ excited by the left-handed radiation ($E_g^{opt}$(A$_+$) = Δ − $\lambda_l$ + $\lambda_u$ + $m_l$ − $m_u$) and the one corresponding to the right-handed light ($E_g^{opt}$(B$_-$) = Δ − $\lambda_l$ + $\lambda_u$ − $m_l$ + $m_u$) split by the magnitude of 2|$m_l$ − $m_u$|. Amazingly, inversed chirality of the incident light sees different $E_g^{opt}$ in the valley polarized system, indicating the possibility to judge the valley polarization utilizing noncontact and nondestructive circularly polarized optical means.

Above discussions establish the general rule to hunt for ferrovalley materials with spontaneous valley polarization, that is the coexistence of the SOC effect with the intrinsic exchange interaction. Here, following the tactic, we predict a certain material: 2H-VSe$_2$ monolayer. As a peculiar ferromagnetic semiconductor [33, 34] among TMDs, it possesses intrinsic magnetic moment with the magnitude of 1.01 $\mu_B$ in the V-3$d$ orbitals, implying remarkable exchange interaction, and then significant spontaneous valley polarization. More excitingly, on the basis of mean field theory and Heisenberg model, its estimated Curie temperature reaches up to ∼ 590 K, in accordance with Pan's work [34], declaring that it could be used in valleytronics well above room temperature.

The calculations of monolayer VSe$_2$ are performed within density-functional theory (DFT) using the accurate full-potential projector augmented wave (PAW) method, as implemented in the Vienna *ab initio* Simulation Package (VASP) [35]. The exchange-correlation potential is treated in Perdew-Burke-Ernzerhof (PBE) form [36] of the generalized gradient approximation (GGA) with a kinetic-energy cutoff of 600 eV. The convergence criterion for the electronic energy is 10$^{-6}$ eV and the structures are relaxed until the Hellmann-Feynman forces on each atoms are less than 1 meV/Å. For the optical property calculations, we adopt our own code OPTICPACK, which has been successfully applied to study the spin-dependent optical properties in ferromagnetic materials [37]. The imaginary part of the complex dielectric function $\varepsilon_2$ excited by circularly-polarized light is calculated using the following relations:

$$[\varepsilon_2]_{\pm}^{\uparrow(\downarrow)}(E) = \frac{4\pi^2}{\Omega} \sum_k W_k \sum_{v,c} |p_{\pm}^{\uparrow(\downarrow)}|^2 \frac{\delta(E_c^{\uparrow(\downarrow)} - E_v^{\uparrow(\downarrow)} - E)}{(E_c^{\uparrow(\downarrow)} - E_v^{\uparrow(\downarrow)})^2}, \quad (5)$$

here $p_{\pm}^{\uparrow(\downarrow)}(k) = (p_x^{\uparrow(\downarrow)}(k) \pm ip_y^{\uparrow(\downarrow)}(k))/\sqrt{2} = <\psi_{c,k}^{\uparrow(\downarrow)}|p_{\pm}|\psi_{v,k}^{\uparrow(\downarrow)}>$ is the electron momentum matrix element of circular polarization between the VB states (v) and the CB states (c). The momentum operator defines as $p = -im_e[r, H]/\hbar$. Noted that the SOC term has been included in the effective one-electron Hamiltonian. $E$ is the photon energy, and $\Omega$ is the cell volume. Spin-flip effects due to SOC are tiny and therefore neglected here [38].

When we ignore the magnetism in monolayer $VSe_2$, as shown in Fig. 2(a), the band structure is essentially similar to the representative one for TMDs (Fig. 1b), except that it is a metal with the Fermi level passing through the states predominantly comprised of $d_{x^2-y^2}$ and $d_{xy}$ orbitals on cation-V. Thankfully, the intrinsic exchange interaction of unpaired $d$ electrons, equivalent to a tremendous magnetic field $\sim 1.59 \times 10^4$ T, completely splits the degenerated spin-up and spin-down components of the states occupied near the fermi level (see Fig. 2b). As a result, the system manifests ferromagnetic semiconductor with a narrow indirect band gap. Though the top of the VB is located in the $\Gamma$ point, the direct band gap remains at two valleys. The relatively small (compared with group-VI dichalcogenides) but non-negligible SOC effect, combined with the strong exchange interaction originating from intrinsic magnetic moment of V-3$d$ electrons induce valley polarization, as shown in Fig. 2c and Fig. 2d.

When the magnetic moment is positive (Fig. 2c), the spin splitting of states mainly occupied by $d_{x^2-y^2}$ and $d_{xy}$ orbitals equals to $|2m_l - 2\lambda_l| \sim 0.85$ eV in the valley $K_+$, which is much smaller as $|2m_l + 2\lambda_d| \sim 1.01$ eV in the valley $K_-$. Reversely, that of primarily $d_{z^2}$ states is with a relatively greater value at the point $K_+$ ($|2m_u - 2\lambda_u| \sim 1.12$ eV) than at $K_-$ ($|2m_u + 2\lambda_u| \sim 1.10$ eV), due to the opposite sign between $\lambda_u$ and $m_u$. By means of these key parameters and some others, we compare the band structures received from the Hamiltonian we constructed in Formula (1) with the DFT results. The excellent agreement in Fig.4a warrants the validity of the two-band **k·p** model we adopted here to describe the electronic properties of valley-polarized TMDs monolayers close to the valleys.

More importantly, we analyze the band gaps and find that it is smaller at valley $K_+$ than at $K_-$ with energy difference $|2\lambda_l - 2\lambda_u| \sim 0.09$ eV, which will directly reflect in the optical properties excited by circularly polarized light. Compared with the $E_g^{opt}$ related to the left-handed radiation, the right-handed one experiences a blue-shift (Fig. 3a). When the magnetic moment is inverted, as clearly displayed in Fig. 2d, our interested valley polarization possess reversed polarity, which causes the red shift of $E_g^{opt}$ excited by right-handed light, in comparison to the left-handed one (Fig. 3b).

Now that we have revealed the spontaneous valley polarization in monolayer $VSe_2$, it is also interesting to inspect the Berry curvature, which has crucial influence on the electronic transport properties and is the kernel parameter to various Hall effects. Here, we consider the spin-resolved nonzero $z$-component Berry curvature from the Kubo-formula derivation [39]:

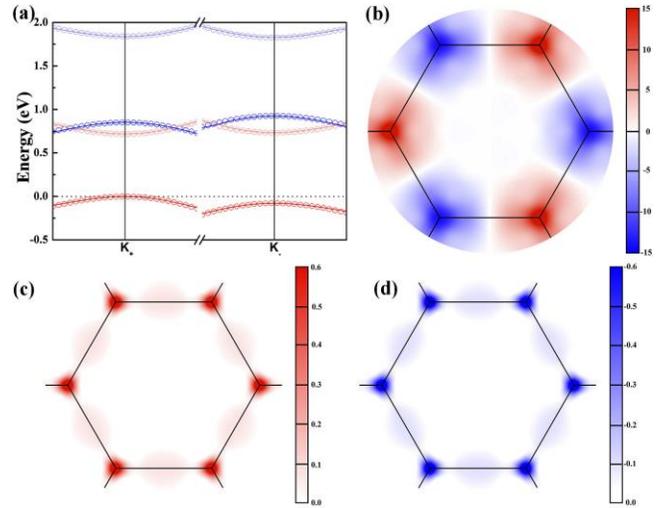

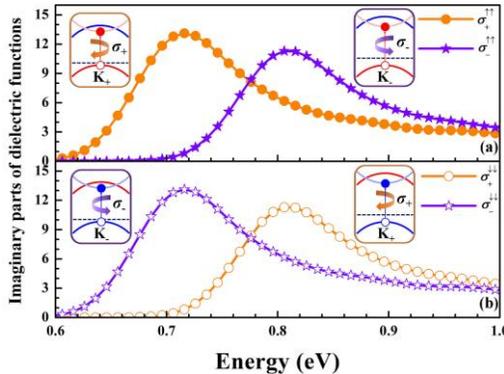

FIG. 3 (color online). The imaginary parts of complex dielectric function $\varepsilon_2$ excited by left-handed radiation $\sigma_+$ and right-handed radiation $\sigma_-$ for monolayer $VSe_2$ with (a) positive magnetic moment (corresponding to Fig. 2c) and (b) negative one (related to Fig. 2d). Insets are the schematic interband transitions related to certain $E_g^{opt}$.

FIG. 4 (color online). (a) Comparison of bands from the two-band **k·p** model (open circles) and the corresponding first-principles results (solid lines). The values of parameters in Hamiltonian are obtained via an optimal fit to the first-principles bands as the following in units of eV: $\Delta = 0.855$, $\varepsilon = 0.851$, $t_{12} = 0.343$, $t_{11}' = 0.171$, $t_{12}' = -0.174$, $t_{22}' = 0.087$, $\lambda_l = 0.038$, $\lambda_u = -0.007$ [40], $m_l = 0.464$, $m_u = 0.555$. (b) Contour maps of berry curvatures in the $k$ space for bands mainly occupied by $d_{x^2-y^2}$ and $d_{xy}$ characters in units of Å$^2$. Summation of Berry curvatures in the point $k$ and its space inversion are shown in (c) and (d) for positive and negative valley polarization cases, accordingly.

$$\Omega_{n,z}^{\uparrow(\downarrow)}(\boldsymbol{k}) = -\sum_{n'\neq n} \frac{2\operatorname{Im}\langle\varphi_{n,\boldsymbol{k}}^{\uparrow(\downarrow)}|v_x|\varphi_{n',\boldsymbol{k}}^{\uparrow(\downarrow)}\rangle\langle\varphi_{n',\boldsymbol{k}}^{\uparrow(\downarrow)}|v_y|\varphi_{n,\boldsymbol{k}}^{\uparrow(\downarrow)}\rangle}{(E_{n'}^{\uparrow(\downarrow)} - E_n^{\uparrow(\downarrow)})^2}, \quad (6)$$

where $v$ is velocity operator. The berry curvatures ($\Omega_{l,z}(\boldsymbol{k}) = \Omega_{l,z}^{\uparrow}(\boldsymbol{k}) + \Omega_{l,z}^{\downarrow}(\boldsymbol{k})$) for the bands with major contribution from $d_{x^2-y^2}$ and $d_{xy}$ states of V atoms, i.e. the summation between blue and red ones in Fig. 2, have been calculated. For a system with equilibrium valleys, $\Omega_{l,z}(\boldsymbol{k})$ is an odd function in the momentum space due to time-reversal symmetry and broken space inversion symmetry. Although the absolute values in opposite valleys are no longer identical in the ferrovalley material, Berry curvatures still have opposite sign, as displayed in Fig. 4b. We would like to emphasize that reversal of valley polarization makes the absolute values of Berry curvatures in valleys $K_+$ and $K_-$ exchanged. The sign of Berry curvature, however, stay the same. To explore the difference between $\Omega_{l,z}(k)$ and $\Omega_{l,z}(-k)$, they are summated. For the case with positive valley polarization induced by positive magnetic moment (Fig. 4c), the absolute value of Berry curvature in $K_+$ valley is greater than the one in the valley $K_-$, giving rise to a positive summation. Not surprisingly, same value with opposite sign is obtained when the valley polarization has been inversed (Fig. 4d). Consequently, Berry curvatures, as circularly polarized radiations, are another effective methods to determine the occurrence of valley polarization and its polarity reversal.

As we know, a direct result related to the sign change of Berry curvatures in different valleys is a new form of Hall effect, namely valley Hall effect, which has been widely investigated in systems with two-dimensional honeycomb lattice [4, 6, 9, 41-43]. Due to the coexistence of spin and valley Hall current, long-lived spin and valley accumulations on sample sides brings charming phenomena, such as emission of photons with opposite circular polarizations on the two boundaries, and provides a route toward the integration of spintronics and valleytronics [9].

We point out that the valley Hall effect in ferrovalley materials possesses a more interesting feature, i.e. the presence of additional charge Hall current deriving from the spontaneous valley polarization. Analogous to the anomalous Hall effect in ferromagnetic materials, we name this effect in ferrovalley materials as *anomalous valley Hall effect*.

Since the charge Hall current is undoubtedly the simplest one to be experimentally measured, the anomalous valley Hall effect offers a possible way to realize data storage utilizing ferrovalley materials. An example for moderate hole doping VSe$_2$ with Fermi energy lying between the VB tops of $K_+$ and $K_-$ valleys is displayed in Fig. 5. It is intriguing to point out that the *p*-type VSe$_2$ possesses 100% spin-polarizability. Considering the almost zero Berry curvature near the center of Brillouin zone, we assume in advance that the carries from the $\Gamma$ point and its neighbors pass through the ribbon directly without transverse deflection. In addition, skew-scattering and other effects due to intervalley scattering are ignored here [4]. When the *p*-type VSe$_2$ possess positive valley polarization, the majority carriers, i.e. spin-down holes from $K_+$ valley, gain transverse velocities toward left side in the presence of external electric field. The accumulation of holes in the left boundary of the ribbon generate a charge Hall current which can be detected as a positive voltage. When the polarity of valley reversed, spin-up holes from $K_-$ valley, as net carriers, accumulate in the right side of the sample due to the negative Berry curvature and then lead to measurable transverse voltage with opposite sign. Noted that in the anomalous valley Hall effect, there exists only one type of carrier coming from a single valley, resulting in the combination of valley, spin and charge Hall current.

Based on the anomalous valley Hall effect, the electrically reading and magnetically writing memory devices are coming up. The binary information is stored by the valley polarization of the ferrovalley material, which could be controlled by the magnetic moment through an external magnetic field. And it can be easily "read out" utilizing the sign of the transverse Hall voltage. Besides the nonvolatile data storage, the ferrovalley materials with spontaneous large valley polarization are ideal candidates for valley filter, valley valve and other promising valleytronic devices [5, 28, 44].

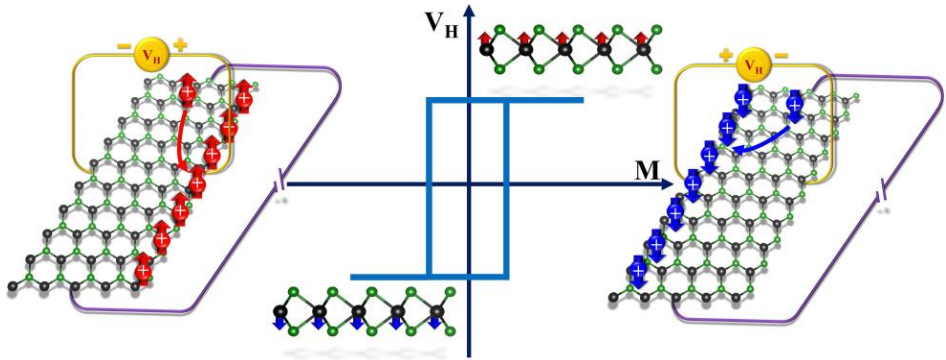

FIG. 5 (color online). Sketch of data storage utilizing hole-doped ferrovalley materials based on anomalous valley Hall effect. The carriers denoted by white '+' symbol are holes. Upward arrows in red color and downward arrows in blue color represent spin-up and spin-down carriers, respectively.

To summarize, we introduce a new concept, i.e. ferrovalley material in our work. As a new ferroic-family number, its potential coupling with ferroelectric, ferromagnetic, ferroelastic and ferrotoroidic properties may provide novel physics in multiferroic field and promote technological innovation. Taking monolayer 2H-VSe$_2$ as an example, we reveal the chirality-dependent optical band gap in it. More interestingly, such system could demonstrate *anomalous valley Hall effect*, indicating the potential use of ferrovalley materials in nonvolatile data storage and other valleytronic devices. We strongly advocate experimental efforts on monolayer 2H-VSe$_2$ and other 2H-phase V-group dichalcogenides, where a series of ferrovalley materials are very likely to hide. It is of great importance in paving the way to the practical applications of valleytronics.

This work was supported by the National Key Project for Basic Research of China (Grant Nos. 2014CB921104 and 2013CB922301), the National Natural Science Foundation of China (Grant Nos. 51572085 and 11525417), ECNU Outstanding Doctoral Dissertation Cultivation Plan of Action (No. PY2015048). Computations were performed at the ECNU computing center. We sincerely acknowledge useful discussions with Prof. J. Feng and Prof. F. Zhang.

*cgduan@clpm.ecnu.edu.cn.